\documentclass[%
reprint,
unsortedaddress,
showpacs,
footinbib,
 amsmath,amssymb,
 aps,
]
{revtex4-1} 

\usepackage{graphicx}
\usepackage{dcolumn}
\usepackage{bm}

\begin{document}

\preprint{APS/123-QED}

\title{Deformation in $^{28}$Si$^*$ produced via $^{16}$O + $^{12}$C reaction  }

\author{S. Kundu}
\email {skundu@vecc.gov.in}
\author{C. Bhattacharya}
\author{S. Bhattacharya}
\author{T. K. Rana}
\author{K. Banerjee}
\author{S. Muhkopadhayay}
 \author{D. Gupta}
 \altaffiliation{Present address: Dept. of Physics and Centre for Astroparticle Physics and Space Science, Bose Institute,  Block EN, Sector V, Bidhan Nagar, Kolkata - 700 091, India}
\author{ A. Dey}
\author{ R. Saha}
\affiliation{Variable Energy Cyclotron Centre, 1/AF, Bidhan Nagar, Kolkata - 700 064, India}

\begin{abstract}
The  energy spectra of  the $\alpha$ particles emitted in the reactions $^{16}$O~(7-10 MeV/nucleon)~+~$^{12}$C have been measured in the center of mass angular range of 25$^\circ$ $\lesssim \theta_{c.m.} \lesssim$ 70$^\circ$.  The experimental energy spectra have been compared with  those obtained from the statistical model calculation  with  ``deformability'' parameters predicted by rotating liquid drop model (RLDM) and also fitted the same with optimized ``deformability'' parameters,  which are quite different from the respective RLDM values. The data have also been found to be explained quite well using  ``frozen'' deformation approximation, where the ``deformability'' parameters have been kept fixed at RLDM values of the parent nucleus throughout the decay process. The effective radius in the latter case is smaller compared to that obtained using the optimized parameters; however, in both cases, the deformations (effective radii) are larger than the corresponding RLDM values.  So, considering the uncertainties in the estimation of  actual compound nucleus deformations, it can, only qualitatively, be said that equilibrium orbiting, which is similar to particle evaporation in time scale, could also be one of the contributing factors for the observed deformation.
\end{abstract}

\maketitle
{\setlength{\baselineskip}%
{1\baselineskip}

\section{\label{sec1:level}Introduction}

Exploring the role of clustering in  fragment emission mechanism is a  subject of current interest in low energy nuclear reaction. Several attempts have recently been made  to understand the effect of cluster structure on the   reaction mechanism of light $\alpha$-cluster nuclei, e.g., $^{20}$Ne + $^{12}$C \cite{Shapira1,Shapira2,Chandana3,Aparajita1}, $^{24}$Mg + $^{12}$C \cite{Dunnweber1}, $^{28}$Si + $^{12}$C  \cite{Shapira3, Shapira4} etc. In each case, an enhancement in the yield and/or resonance-like excitation function in a few outgoing channels  not very different from the entrance channel has been observed,  which was indicative of significant contribution  from the deep-inelastic orbiting (DIO) mechanism. Through this process, a long-lived  dinuclear composite is formed \cite{Scanders}, which is not fully equilibrated in all (in particular, shape) degrees of freedom and therefore decays preferentially around the entrance channel; at higher excitation energy, however, the decay is expected to be of fully equilibrated in nature. It is thus interesting to probe into the limit of survival of such dinuclear composites at higher excitations.

The study of deformation of the excited composite is  important  to differentiate the DIO composite from the fully equilibrated compound nucleus (CN) because the deformation should be larger in the former case. Detailed studies of the deformation of $^{40}$Ca$^{*}$ \cite{Forna2,Rousseau1} and $^{32}$S$^{*}$ \cite{Aparajita2}   were done  recently with the help of  light charged particle (LCP) spectroscopy to extract ``anomalous'' deformation in these cases confirming the presence of DIO mechanism.  For these systems,  orbiting ($^{40}$Ca$^{*}$  \cite{Shapira3, Shapira4}, $^{32}$S$^{*}$ \cite{Shapira1,Shapira2,Chandana3})had already been conjectured from the fragment emission studies. Recently, we have reported that, for $\alpha$-cluster system $^{16}$O + $^{12}$C \cite{samir}, there was enhancement in the boron product yield at excitation energies $\sim$67-85 MeV (beam energy in the range of 7-10 MeV/nucleon), which indicated the survival of long-lived  dinuclear orbiting composites at such high excitation energies. An orbiting dinuclear complex is assumed to be more deformed than a fully shape equilibrated CN; this motivated us to look for such enhanced deformation in a $^{16}$O + $^{12}$C dinuclear system at same excitation energies  using the LCP spectroscopy technique. In this technique, LCPs, evaporated from the excited composites, are used to study the properties of hot rotating  nuclear systems as a function of excitation energy, angular momentum, and deformation \cite{Chandana2,Govil1,Choudhury1, Rana1,Rana2, Fornal1,Fornal2, Aparajita2,Chandana1,Dipa1,Dipa2,Parkar1,Ajitanand1,Kildir1} using  statistical model codes  e.g., CASCADE}\cite{Puhlh1}, GANES \cite{Ajitanand2}, CACARIZO \cite{Viesti1}.

The article has been arranged as follows. The experimental setup has been described in Sec.~\ref{sec2:level}. The experimental results and analysis have been  presented in Sec.~\ref{sec3:level}. The details of the statistical model calculation has been given in Sec.~\ref{sec4:level}. The extracted values of the deformation of $^{28}$Si$^*$ and their possible implications have been discussed in Sec.~\ref{sec4_5:Discussion}. Finally, the summary and conclusion have been given in Sec.~\ref{sec5:level}.

\section{\label{sec2:level}Experimental details}
The experiment was performed at the Variable Energy Cyclotron  Centre, Kolkata, using $^{16}$O ion beams  of energies  117, 125, 145 and 160 MeV. Self-supporting $^{12}$C of thickness $\sim$ 514 $\mu$g/cm$^{2}$ was used as a target. The $\alpha$ particles  were detected using a Si(SB) telescope ($\sim$ 10$\mu$m $\Delta$E, $\sim$ 5 mm E). The typical solid angle covered by the telescope was $\sim$ 0.3 msr. The calibration of the telescope was done  using elastically scattered $^{16}$O ion from a Au target, as well as a $^{229}$Th-$ \alpha $ source. Inclusive energy distributions for the LCPs were measured in the laboratory angular range of 11$^{0}$ - 29$^{0}$, which covered the angular range $\sim$ 25$^{0}$ - 70$^{0}$ in the center-of-mass frame.

\section{\label{sec3:level}Experimental results}
\subsection{\label{sec3a:level} Energy spectra }
\begin{figure}
\includegraphics[scale=0.42]{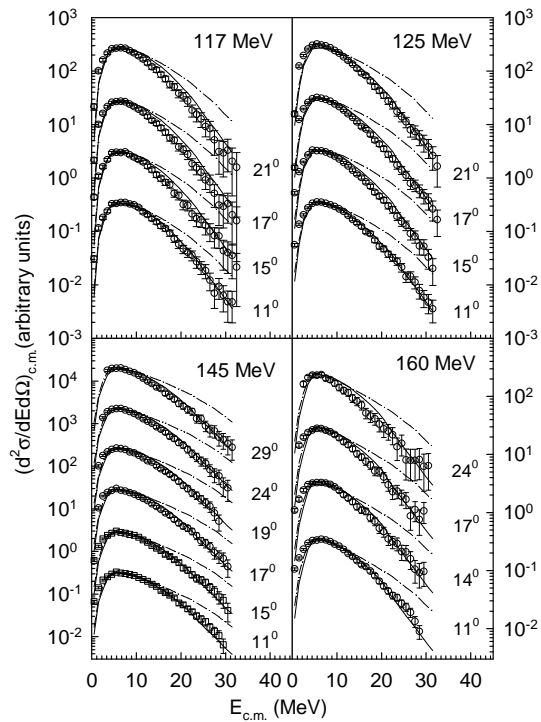}
\caption{\label{fig1:epsart} Energy spectra (c.m.) of $\alpha$ particles obtained at different  angles for different beam energies. The symbols represent  experimental data.  The dash-dot-dashed and solid lines  represent  CASCADE calculations with RLDM  and   optimized values of spin dependent `deformability' parameters, respectively (see Table ~\ref{tab2:table}). All  experimental  and calculated spectra, starting from the lowest angle, were multiplied by 10$^{-2}$, 10$^{-1}$, 10$^{0}$, 10$^{1}$, 10$^{2}$, 10$^{3}$, respectively.}
\end{figure}

\begin{figure}
\includegraphics[scale=0.4]{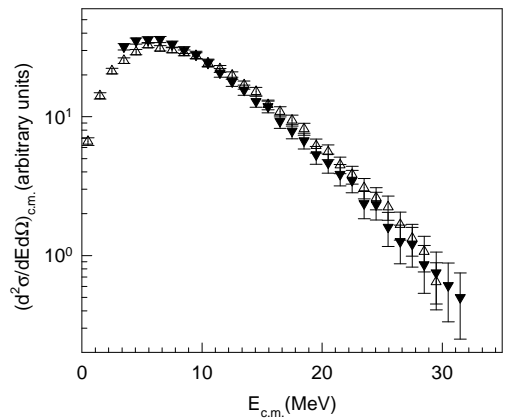}
\caption{\label{fig2:epsart} Comparison of the energy spectra  (c.m.) of $\alpha$-particles obtained at  c.m. angles 26$^{0}$(open triangles) and 66$^{0}$ (inverted triangles) for beam energy 145 MeV. }
\end{figure}

The  center-of-mass (c.m.) energy spectra of the   $\alpha$ particles  at different beam energies are shown in Fig.~\ref{fig1:epsart}. Because of the inclusive nature of the spectra, there may be some admixture of contributions of other direct reaction mechanism (such as pre-equilibrium  emission, etc.) to the equilibrium   emission  spectra at forward angles in particular. In Fig.~\ref{fig2:epsart}, the   energy  spectra obtained at lab angles  11$^{0}$($\theta_{c.m.}$= 26$^{0}$, open triangles) and 29$^{0}$($\theta_{c.m.}$= 66$^{0}$, solid inverted triangles) at $E_{lab}$=145 MeV are compared; it is seen that the slopes of the both spectra are completely matched with each other. It is thus clear that the effects of other direct reaction mechanisms are not significant at the measured angles.
\hspace{2.0cm}

\subsection{\label{sec3b:level} Angular distribution }
\begin{figure}
\includegraphics[scale=0.482]{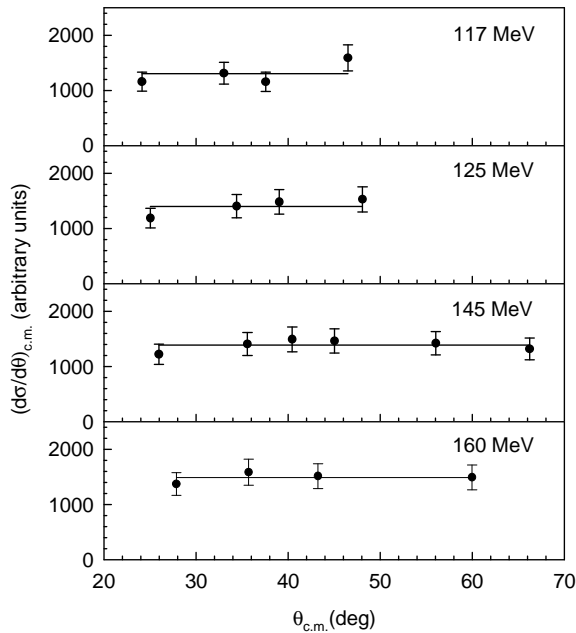}
\caption{\label{fig3:epsart} Angular distribution of $\alpha$ particles as a function of c.m. angles, $\theta_{c.m.}$. Symbols represents the experimental data and solid lines show fit to the data obtained using  (d$\sigma$/d$\theta$)$_{c.m.}$ = constant.}
\end{figure}

The c.m. angular distributions, (d$\sigma$/d$\theta$)$_{c.m.}$, obtained by integration of  the  c.m. energy distributions for the beam energies of 117, 125, 145 and 160 MeV, are shown in Fig.~\ref{fig3:epsart} as a function of $\theta_{c.m.}$. In all cases, the  values of (d$\sigma$/d$\theta$)$_{c.m.}$  are found to be constant over the whole range of observed c.m. angles. So, (d$\sigma$/d$\Omega$)$_{c.m.}$ is $\propto$  1/sin{$\theta_{c.m.}$},  which is  characteristic of the emission from an equilibrated composite nucleus.

\subsection{\label{sec3c:level} Average velocity }
\begin{figure}
\includegraphics[scale=0.475]{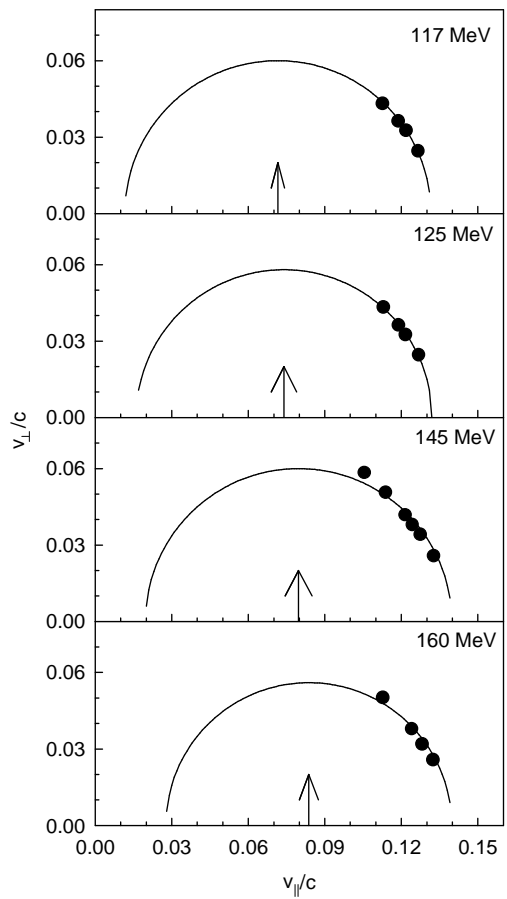}
\caption{\label{fig4:epsart} Average velocity curve. Solid circles correspond to experimental data and solid lines show fit to the data obtained using the equation \textit{v}$_\bot^{2}$ = $(v^{c.m.}_{av})^{2}$ - ($v_{||}$ - $v_{CN}$)$^{2}$ where $v^{c.m.}_{av}$ is the average velocity of $\alpha$-particles in the c.m. frame. The arrow indicates the position of $v_{CN}$.}
\end{figure}

The average velocities, $v_{av}, $ of the   $\alpha$-particles emitted at different angles have been extracted from the average energy, $E_{av}$, which has been calculated using the  expression
\begin{equation}
\label{eq:V_av}
E_{av}=\frac{\sum_{i} E_{i}(d^{2}\sigma/dEd\Omega)_{i}}{\sum _{i}(d^{2}\sigma/dEd\Omega)_{i}},
\end{equation}
where the index \textit{i} covers the whole energy spectrum in the laboratory. The parallel ($v_{||}$) and perpendicular (\textit{v}$_\bot$) components (with respect to the beam direction) of   $v_{av}$ at each angle are plotted in Fig.~\ref{fig4:epsart}. It is seen that they  fall on a circle with the center at CN velocity, $v_{CN}$, and radius of average velocity in c.m., $v^{c.m.}_{av}$,  which implies that the   average velocities (as well as energies) of the  $\alpha$ particles are independent of the c.m. emission angles. It again indicates that the  $\alpha$ particles are emitted from a fully energy equilibrated source moving with the velocity $v_{CN}$.

\section{\label{sec4:level}Statistical model calculations}
It is confirmed from the experimental results  that the $\alpha$-particles are emitted from a fully  energy equilibrated composite which may be a CN. So the statistical model code CASCADE \cite{Puhlh1} has been used to explain the present experimental data because this code is based on the assumption that the evaporation of particles takes place from an excited CN which is in full equilibrium with respect to all degrees of freedom.

\subsection{\label{sec4a:level} CASCADE calculations procedure}
The   spin distribution in CASCADE  is usually derived from the known fusion cross section using  the strong-absorption model wherever data are available; otherwise, it is calculated theoretically. The decay of the CN by particle emission is  calculated in the framework of Hauser-Feshbach formalism \cite{Hauser1,Puhlh1}. The probability that the parent nucleus (excitation energy $E_{1}$, spin $J_{1}$, parity $\pi_{1}$)  emits a particle $x$ with an orbital angular momentum $l$, kinetic energy $\epsilon_{x}$ ( $\epsilon_{x}$=$E_{1}$-$E_{2}$ ), and spin $s$ is given by
\begin{equation}
\label{eq:Decay_prob}
\begin{split}
P_{x}d\epsilon_{x}&=\frac{1}{\hbar}\Gamma\left(\epsilon_{x}\right)
=\frac{\rho_{2}(E_{2},J_{2},\pi_{2})}{2\pi\hbar\rho_{1}(E_{1},J_{1},\pi_{1})}\\
&\times   \sum^{J_{2}+s}_{S=\left|J_{2}-s\right|}\sum^{J_{1}+S}_{\substack{{l=\left|J_{1}-S\right|}\\{\left[\pi\right]}}}T_{l}(\epsilon_{x})d\epsilon_{x},
\end{split}
\end{equation}
where  $E_{2}$,  $J_{2}$ and   $\pi_{2}$ are the excitation energy, spin, and parity of the daughter nucleus, respectively, $\rho_{1}$ and $\rho_{2}$ are the level-densities of the excited parent nucleus and daughter nucleus, respectively, and,  \textit{S = $J_{2}$} + \textit{s} is the channel spin. The transmission coefficients $T_{l}$($\epsilon_{x}$) for the scattering of particle ${x}$ on the daughter nucleus (inverse process) are obtained by using standard optical model (OM) potential for elastic scattering. In the OM transmission coefficient calculation,  the   parameters have been taken from the references given in the  Table~\ref{tab1:table}.
\begin{table}
\caption{\label{tab1:table}Input parameters used  for CASCADE calculations for the reactions  $^{16}$O + $^{12}$C at beam energies 117, 125, 145, and 160 MeV.\\}
\begin{ruledtabular}
\begin{tabular}{p{8cm}}

\\

\underline{Angular momentum distribution:} \\
\\
Critical angular momentum $J_{cr}$ as in Table \ref{tab2:table} \\
Diffuseness parameter $\Delta J$ = 1$\hbar$ \\
\\

\underline{OM potentials of the emitted LCP and neutrons} \\
\\
(1) Neutrons: Wilmore and Hodgson \cite{Wilmore1}.\\
(2) Protons: Perey and Perey \cite{Pere1}.\\
(3) $\alpha$-particles: Huizenga and Igo \cite{Hui2}.\\
(4) Multiplication factor of the OM radius: RFACT = 1 \\
\\

\underline{Level-density parameters at low excitation: $E\leq$ 7.5 MeV}\\
\\
(1) Fermi-gas level-density formula with empirical parameters
from Dilg \textit{et al}. \cite{Dilg1}.\\

\\

\underline{Level-density parameters at high excitation: $E\geq$ 15MeV} \\

\\
(1) Fermi-gas level-density formula with parameters from \\
LDM (Myers and Swiatecki \cite{Mayer1})\\
(2) Level-density parameter \textit{a} = \textit{A}/8 MeV$^{-1}$\\
\\

\underline{Yrast line}\\
\\

$I_{eff}$ = $I_{0}$(1+$\delta_{1}$ $J^2$+$\delta_{2}$ $J^4$),
$\delta_{1}$, $\delta_{2}$  are given in Table \ref{tab2:table}.\\
\\
\underline{$\gamma$-ray width (Weisskopf units)}\\
\\
(1) \textit{E1} = 0.001\\
(2) \textit{M1} = 0.01\\
(3) \textit{E2} = 5.0\\
\end{tabular}
\end{ruledtabular}
\end{table}
\subsection{\label{sec4b:level} Input parameters}
The  standard form of CASCADE is quite successful in explaining the   LCP evaporation in the light ion induced reaction in general, where the CN is assumed to be nearly spherical. However, in case of heavy-ion-induced reaction, there is appreciable deviation between the experimental and the predicted LCP evaporation spectra. This deviation is attributed to the deformation of the excited compound system which is angular momentum dependent. Therefore, to explain the LCP energy spectra, the effects of the deformation of the CN should be included in the statistical model calculations.  The deformation affects the particle spectra in two ways. First, it lowers the effective emission barrier, and second, it increases the moment of inertia. The first effect modifies the transmission coefficients for the evaporated particles which may be   taken care  of by increasing the radius parameter of OM potential. However, the change in moment of inertia affects the level density and the slope of the particle spectrum. This can be taken care of  by incorporating    the spin-dependent deformability parameters \cite{Hui1,Govil1,Dipa2}. For level density calculations, excited energies have been divided into three regions.

Region I (low excitation energy, \textit{E} $\le$ 3 to 4 MeV): Here, the experimentally known discrete levels are used for all the    nuclei produced in the cascade. In some cases, known high-spin states at higher excitation energy are included as yrast levels in region II.

Region II (medium excitation energy, 4 MeV $\le$ \textit{E} $\le$ 7.5 MeV) : Analytical level density formula is used in this region. The parameters ${a}$ and $\Delta$ are  deduced empirically for each nucleus from the work of Vonach \textit{et al.} \cite{Vonach1} and Dilg \textit{et al}. \cite{Dilg1}. The excitation energy is corrected for the parity effects.

Region III (high excitation energy, \textit{E} $\ge$ E$_{LDM}$): Shell effects and parity corrections are neglected in this region. The same formula is then used but with LDM parameters taken from Ref. \cite{Mayer1}.

Between regions II and III, the level density parameters are interpolated linearly. The parameters are given in Table \ref{tab1:table}.  The level density used in regions II and III for a given angular momentum \textit{J}  and excitation energy \textit{E}  is given by well-known Fermi-gas expression \cite{Aparajita2} with equidistant single-particle levels and a constant level density parameter $a$:
\begin{eqnarray}
\label{eq:Level_density}
\rho(E,J)&=&\frac{(2J+1)}{12}a^{1/2}\left(\frac{\hbar^2}{2\textsl{I}_{eff}}\right)^{3/2}\frac{1}{(E+T-\Delta-E_{J})^{2}} \nonumber
\\
&&\times exp\left[2\{a(E-\Delta-E_{J})\}^{1/2}\right],
\end{eqnarray}
where  \textit{T} is the thermodynamic temperature and $\Delta$ is the pairing correction. The rotational energy, $E_{J}$, is expressed as,
\begin{equation}
\label{eq:E_rot}
E_{J}= \left(\frac{\hbar^2}{2\textsl{I}_{eff}}\right)J(J+1).
\end{equation}
The effective moment of inertia, $I_{eff}$, is written as,
\begin{equation}
\label{eq:I_eff}
I_{eff}= I_{0}(1+\delta_{1}J^2+\delta_{2}J^4),
\end{equation}
where \textit{I$_{0}$} ($ =\frac{2}{5}A^{5/3}r_{0}^2 $) is the  rigid-body moment of inertia, $\delta_{1}$ and $\delta_{2}$ are deformability parameters, \textit{r$_{0}$} is radius parameter, and  $\textit{a}$ is the level density parameter. So, from  the above equations, it is clear that by changing \textit{r$_{0}$}, \textit{a}, $\delta_{1}$, and $\delta_{2}$, it may be possible to reproduce the experimental spectra.
By increasing \textit{r$_{0}$},  both transmission coefficient and level  density  will be  affected. It reduces the potential barrier, leading to the increase of transmission coefficient. Simultaneously, the increase in \textit{r$_{0}$} leads to  increase of the \textit{I$_{eff}$} and \textit{vis-\`a-vis} the available phase space. We have chosen \textit{r$_{0}$} = 1.29 which reproduced the lower energy part of the spectra. A similar value was used in Ref. \cite{Mahboub1}. The level density parameter $\textit{a}$ was taken to be \textit{A}/8 as in previous works \cite{Hui1,Chandana2}.To reproduce the present  experimental spectra, we have only changed the deformability parameters $\delta_{1}$ and $\delta_{2}$, as in the previous works \cite{Govil1,Viesti1,Hui1}.

\subsection{\label{sec4c:compare} Experimental spectra and CASCADE predictions}

The measured energy spectra  have been compared with the respective CASCADE \cite{Puhlh1} calculations which  have been shown in Fig.1. The  critical angular momenta, \textit{J$_{cr}$}, used in the calculation, are 20, 21, 22, and 23($\hbar$) for  the bombarding energies of 117, 125, 145, and 160 MeV, respectively \cite{samir}. The other input parameters are given in Table~\ref{tab1:table}. The dash-dot-dashed lines represent the results of the CASCADE calculation with the radius parameter \textit{r$_{0}$} = 1.29 and the default values of  the deformability parameters, $\delta_{1}^{A}$ = 3.7$\times$ 10$^{-4}$ and $\delta_{2}^{A}$ = 1.1$\times$ 10$^{-6}$, at all beam energies  which have been predicted by rotating liquid drop model \cite{Cohen1}. The low-energy parts of the experimental $\alpha$ particle  spectra are found to  match with the theoretical spectra, but higher energy parts do not. To reproduce the whole spectra, we have followed the procedure proposed by Huizenga \textit{et al.} \cite{Hui1}. The deformability parameters $\delta_{1}$ and $\delta_{2}$ have been  suitably optimized  to reproduce the experimental spectra, which in effect modified    the phase space for statistical decay by relocation of the yrast line. The results with these modified deformability parameters, are shown by solid lines in Fig. 1 and the optimized  values of deformability parameters, $\delta_{1}^{B}$ and $\delta_{2}^{B}$, are given in Table ~\ref{tab2:table}.

\begin{table}
\caption{\label{tab2:table} The values  different sets of  `deformability' parameters: A -  obtained from RLDM,  and, B -  obtained by fitting the experimental data (see text). E$_{lab}$, \textit{E}, \textit{J$_{cr}$} and \textit{J$_{av}$} are the beam energy, excitation energy, critical angular momentum and average angular momentum, respectively.\\}
\begin{ruledtabular}

\begin{tabular}{cccccccc}
 $E_{lab}$ & \textit{E} & $J_{cr}$ & $J_{av}$ &  $\delta_{1}^{A}$ & $\delta_{2}^{A}$     & $\delta_{1}^{B}$ & $\delta_{2}^{B}$       \\
\hline
\\[0.1pt]
117 & 67 & 20 & 13 & 3.7 $\times$10$^{-4}$ & 1.1 $\times$10$^{-6}$  & 1.9 $\times$10$^{-3}$ & 2.0 $\times$10$^{-8}$  \\
125 & 70 & 21 & 14 & 3.7 $\times$10$^{-4}$ & 1.1 $\times$10$^{-6}$ &2.1 $\times$10$^{-3}$ & 2.0 $\times$10$^{-8}$ \\
145 & 79 & 22 & 15 & 3.7 $\times$10$^{-4}$ & 1.1 $\times$10$^{-6}$  & 2.3 $\times$10$^{-3}$ & 2.0 $\times$10$^{-8}$ \\
160 & 85 & 23 & 15 & 3.7 $\times$10$^{-4}$ & 1.1 $\times$10$^{-6}$  & 2.5 $\times$10$^{-3}$ & 2.0 $\times$10$^{-8}$  \\
\end{tabular}

\end{ruledtabular}
\end{table}

\section{\label{sec4_5:Discussion}Discussion}

It is apparent from  Table~\ref{tab2:table} that the deformability parameters had to be modified from the corresponding RLDM values considerably to fit the experimental spectra. Following the    empirical procedure given in Refs. \cite{Chandana2,Rousseau1}, it can be shown that the above change in the  deformability parameters is indicative of substantial enhancement of the deformation of the excited composite over the corresponding ground state (RLDM) deformation. Before we look further into the origin of such deformation, it may be worthwhile to see if there is any other alternative explanation of the observed deviation of the experimental spectra from the respective standard statistical model prediction.

Though the present formalism of angular momentum dependent level density is largely successful in  explaining experimental LCP spectra, the magnitude of enhancement required is quite large and lacking proper explanation \cite{Ormand1, Lauritzen1}. So an alternative approach, based on frozen degrees of freedom, has been proposed, which has been shown to reproduce the data quite well  \cite{Fornal3,Fornal4}. In this approach, it is assumed that the deformation of the CN is frozen during the decay, i.e., there is no change of shape of the nascent final nucleus; so the phase space is calculated using RLDM deformation of the parent nucleus, rather than that of the usual daughter nucleus.  This indicates that the dynamical effects such as shape relaxation should be taken into account to properly understand the phenomenon of particle evaporation from a deformed CN.

\begin{figure}
\includegraphics[scale=0.4]{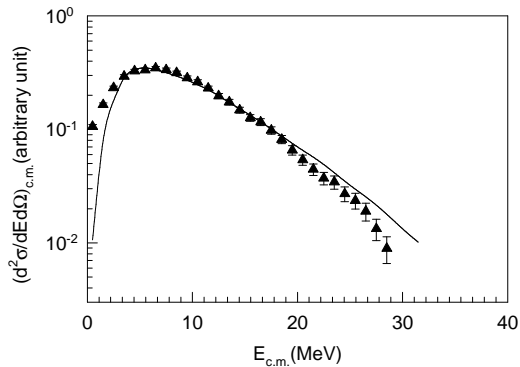}
\caption{\label{fig5:Frozen} Comparison of the experimental energy spectra (c.m.) of $\alpha$-particles  (triangles) with the same obtained by theoretical calculations  for beam energy 160 MeV. Solid  curve represents  the CASCADE calculations done using `frozen' deformation ($I_{eff}/I_{0}$) obtained using  RLDM deformability parameters for $J = 23 \hbar$.}
\end{figure}

The formalism of ``frozen deformation'' has been applied in the present CASCADE calculation. In the case of $\alpha$ particle emission, the effect of frozen deformation on the energy spectrum may be taken into account in the following way. As the $\alpha$ particles are emitted predominantly at the initial stage of the decay cascade, the deformation may be ``frozen'' at its value corresponding to the  highest angular momentum that the CN may have, which is $\approx J_{cr}$. So in the present CASCADE calculation, the deformation has been kept fixed throughout  by freezing the value of $I_{eff}/I_{0}$, which has been calculated using a fixed value of  $J \approx J_{cr}$ using Eq.~\ref{eq:I_eff} with $\delta_{1}$ and $\delta_{2}$ obtained from RLDM (see Table ~\ref{tab2:table}). The results of the CASCADE calculation with frozen deformation  have been shown in Fig.~\ref{fig5:Frozen} along with the experimental data. It is observed that the CASCADE prediction with frozen deformation  \cite{Fornal3,Fornal4} for $J = J_{cr} = 23 \hbar$ (solid curve) is also in fair agreement with the experimental data.

To investigate further into the nature of deformation, we have computed the values of  average effective radius parameter,$\left\langle r_{eff}\right\rangle$, for all sets of deformability parameters (Table ~\ref{tab2:table}). Following \cite{Aparajita2}, the effective radius parameter, \textit{r$_{eff}$}, is defined as $r_{eff} = r_{0}\sqrt{(1 + \delta_{1}J^2 + \delta_{2}J^4)}$, and the average effective radius parameter is calculated using the expression, $\left\langle r_{eff}\right\rangle$ = $\sum_0^{j_{cr}}r_{eff}(2j+1)/\sum_0^{j_{cr}}(2j+1)$. Then the deviation of $\left\langle r_{eff}\right\rangle$  from \textit{r$_{0}$} is  the measure of deformation, and the deviation of the same from the corresponding RLDM value may be considered as an enhancement. The values of \textit{r$_{eff}$} estimated in the case of `frozen' deformation approach  varied between 1.48 (at 117 MeV) and 1.58 (at 160 MeV), which are, though higher than the corresponding RLDM values ($\left\langle r_{eff}\right\rangle ^{RLDM}$ $\sim$ 1.38 - 1.42), quite lower than the values obtained with the best fitted parameter set ($\left\langle r_{eff}\right\rangle ^{fitted}$ $\sim$ 1.52 - 1.66). From the above discussion, it may be said that, though the present study indicates deformation of the excited composite which is higher than the corresponding RLDM value, uncertainty still remains about the actual magnitude of deformation. This limitation of the present study notwithstanding, there is, at least qualitative, indication about some enhanced deformation, which may be linked with orbiting as one of the contributing factors.

\section{\label{sec5:level}Summary and conclusion}
The energy and angular distribution of $\alpha$ particles, emitted in the reactions  $^{16}$O (117, 125, 145 and 160 MeV) + $^{12}$C, have been measured. It has been observed from the angular distribution and average velocity plots that the $\alpha$-particles are emitted from fully energy equilibrated sources, at all beam energies. The measured energy spectra have been compared with the same  predicted by the standard statistical model  calculations. It has been  found that the experimental  $\alpha$ particle energy spectra are properly explained by CASCADE by optimizing the deformability parameters, which are quite different from the respective RLDM values. This appears to be indicative of appreciable change (enhancement) of deformation. However, the understanding does not seem to be so simple and straightforward when one   compares the above with the results of CASCADE calculation with frozen deformation, which has also  been shown to be almost equally effective in explaining the data. In this case, the effective `frozen' deformation turned out to be smaller than that obtained using the optimized parameters. So, the uncertainties about the magnitudes of the actual CN deformations notwithstanding, it can, only qualitatively, be said that equilibrium orbiting, which is similar to particle evaporation in time scale, could also be one of the contributing factors for the observed deformation. However, the present models are too simplistic to predict the actual deformation of the CN; more realistic event-by-event Monte Carlo calculations, taking into account the initial deformation, spin distribution of the CN and their subsequent evolutions, should be performed to have a proper understanding of the CN deformation. In addition,  new experimental inputs (such as measurement of deformation from GDR studies  \cite{Deepak1}) are needed for more comprehensive understanding of the process.

\begin{acknowledgments}
The authors wish to thank the cyclotron operating staff for smooth running of the machine.
\end{acknowledgments}

\end{document}